\documentclass[preprint,superscriptaddress,preprintnumbers,amsmath,amssymb,floats]{revtex4}
\pdfoutput=1
\usepackage{txfonts}
\usepackage{amssymb}
\usepackage{graphicx}
\usepackage{sidecap}
\usepackage{color}

\begin{document}

\title{Predicting Unconventional High Temperature Superconductors in Trigonal Bipyramidal Coordinations}
\author{Jiangping Hu}\email{jphu@iphy.ac.cn }  \affiliation{Beijing National Laboratory for Condensed Matter Physics, and Institute of
Physics, Chinese Academy of Sciences, Beijing 100190, China}
\affiliation{Department of Physics, Purdue University, West Lafayette, Indiana 47907, USA}
\affiliation{Collaborative Innovation Center of Quantum Matter, Beijing, China}

\author{Congcong Le}   \affiliation{Beijing National Laboratory for Condensed Matter Physics, and Institute of
Physics, Chinese Academy of Sciences, Beijing 100190, China}

 \author{Xianxin Wu}   \affiliation{Beijing National Laboratory for Condensed Matter Physics, and Institute of
Physics, Chinese Academy of Sciences, Beijing 100190, China}

\begin{abstract}
\textbf{Cuprates  and iron-based superconductors are two classes of unconventional high $T_c$ superconductors based on 3d transition elements. Recently, two principles, correspondence principle and  magnetic selective pairing rule,  have been emerged to unify their  high $T_c$ superconducting mechanisms.   These principles strongly regulate electronic structures that can host  high $T_c$ superconductivity. Guided by these principles, here we propose high $T_c$ superconducting candidates that are formed by cation-anion trigonal bipyramidal complexes with a $d^{7}$ filling configuration  on the cation ions.   Their superconducting states are expected to  be dominated by  the $d_{xy}\pm id_{x^2-y^2}$ pairing symmetry.
\\
}
\end{abstract}

\maketitle

Almost three decades ago,   cuprates\cite{Bednorz1986}, the $Cu$-based high $T_c$ superconductors, were discovered. Since then, understanding the superconducting mechanism behind unconventional high temperature superconductors  has become a great challenge in condensed matter physics.   In the past six years, new light has been shined to this decades-old problem due to the discovery of iron-based  high $T_c$ superconductors\cite{Kamihara2008-jacs}. The two high temperature superconductors share many common electronic properties\cite{Johnston2010-review}.  In principle,  comparing these two classes of materials,  we may determine the key ingredients that are essential to the  high $T_c$ superconducting mechanism. However,  even if  we have identified them,  without  a realistic  prediction  of new high $T_c$ superconductors, reaching a final consensus will be extremely difficult.

 Most recently,   one of us   emphasized and proposed  two basic principles to unify the understanding for both high $T_c$ superconductors\cite{Hu2015-swave}: (1) the HDDL  correspondence principle, which was first specified in ref.\cite{Huding2012} by Hu and Ding and was generalized   to include other orders later in ref.\cite{DavisLee2013} by Davis and Lee:  the short range magnetic exchange interactions and the Fermi surfaces act collaboratively to achieve high $T_c$ superconductivity and determine pairing symmetries;  (2)  the selective magnetic pairing rule:  the superconductivity is only induced by the magnetic exchange couplings from the superexchange mechanism  through cation-anion-cation chemical bondings but not those from direct exchange couplings resulted from the direct cation's d-d chemical bondings.  These two principles  provide an unified explanation why the d-wave pairing symmetry and the s-wave pairing symmetry are robust respectively in curpates and iron-based superconductors\cite{Hu2015-swave}.  In the meanwhile,  the above two principles can serve as  direct guiding rules to search   high $T_c$ superconductors.  The two principles provide many constrains on electronic structures that can host  high $T_c$ superconductivity.  The detailed summary of these constraints  and their microscopic origins were discussed  in ref.\cite{Hu2015-swave}. Essentially, the two principles suggest that   the electronic environment that  hosts high $T_c$ superconductivity must include quasi-two dimensional bands  formed dominantly by the d-orbitals through a d-p hybridization.   

Here, guided by these principles,  combining with  crystal field theory and   first principle calculations, we predict a new electronic structure that  can host high $T_c$ superconductivity with $d\pm id$ pairing symmetry.

\begin{figure}[t]
\centerline{\includegraphics[height=10 cm]{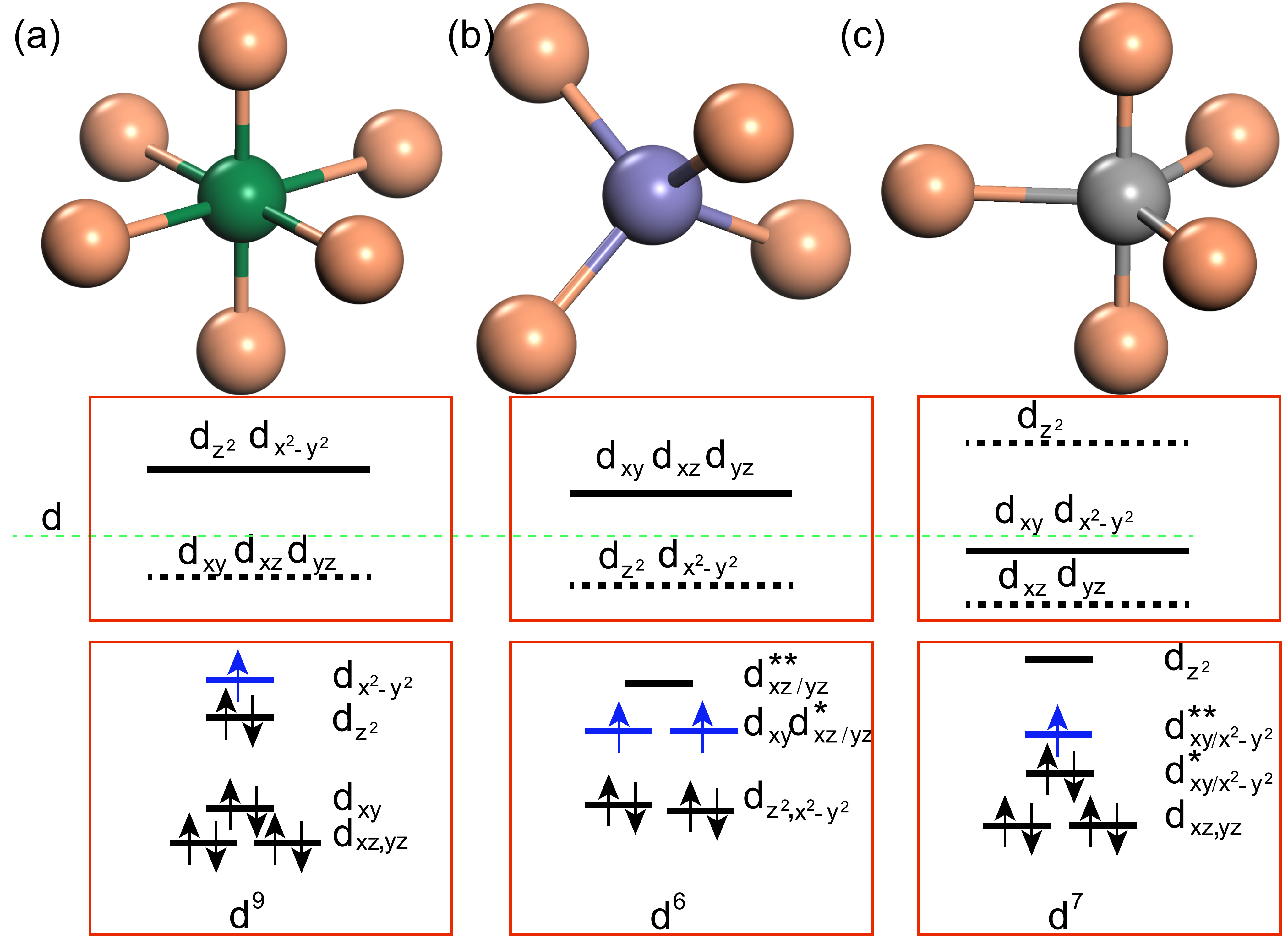}}
\caption{  Structural units, crystal field splitting in one unit complex and  energy splitting in the corresponding two dimensional lattice: (a) Octahedral complex ( cuprates, $CuO_6$ ); (b) Tetrahedral complex (iron-based superconductors, $FeAs_4/Se_4$); (c)  Trigonal bipyramidal complex ($Ni/CoO_3$). 
The $d$ orbitals with the blue color are active ones for superconductivity.
 \label{crystal} }
\end{figure}

We  start to search possible high $T_c$ candidates by analyzing the basic building blocks, namely, the cation-anion complexes. Taking  both cuprates and iron-based superconductors as examples, we check how the principles are satisfied in these two superconductors.  As shown in Fig.\ref{crystal}(a), the Cu atoms in cuprates are in an octahedral complex. In this complex, the five d-orbitals splits into two groups by crystal fields, $t_{2g}$ and $e_g$.  The two orbitals in the $e_g$ group, $d_{z^2}$ and $d_{x^2-y^2}$,  because of their strong couplings to the p-orbitals of the surrounding oxygen atoms, have higher energies. However, only the $d_{x^2-y^2}$ orbitals have strong in-plane couplings to the p-orbitals. Therefore, following  the above rule,  only the electronic band attributed to the $d_{x^2-y^2}$ orbitals can support high $T_c$ superconductivity.  In a two-dimensional layer structure, the $d_{z^2}$ energy level is lowered due to the Jahn-Teller effect and  the $d_{x^2-y^2}$ orbital  is  the single orbital at the highest energy as shown in Fig.\ref{crystal}(a). Thus, it is easy to see that  in this case, an electronic band structure for high $T_c$ superconductors can only be achieved under the $3d^9$($Cu^{2+}$) configuration.   In iron-based superconductors, the Fe atoms are in a tetrahedral  complex. Compared with the  octahedral environment,   the energy levels of  the $t_{2g}$ and $e_g$ orbitals in the tetrahedral complex reverse. The $t_{2g}$ orbitals have higher energy because of their strong couplings to the As/Se anions.  If we further consider two molecular orbitals formed by $d_{xz}$ and $d_{yz}$, one molecular orbital  is strongly coupled to the $e_g$ orbitals and becomes inactive in supporting pairing.  Thus, as shown in Fig.\ref{crystal}(b),   the $3d^6$ ($Fe^{2+}$) configuration is  the  filling level to make the pure $t_{2g}$ orbitals  to dominate electronic band structures close to Fermi energy. The high $T_c$ superconductivity  is thus only achieved under the $3d^{6}$ configuration. From these understandings, we can see that  the two principles  fix the d-orbital filling configuration  if a  structure formed by a given cation-anion complex is a high $T_c$ superconductivity candidate. This result partially explains why  high $T_c$ superconductivity appears to be such a rare phenomena.

\begin{figure}[t]
\centerline{\includegraphics[height=8 cm]{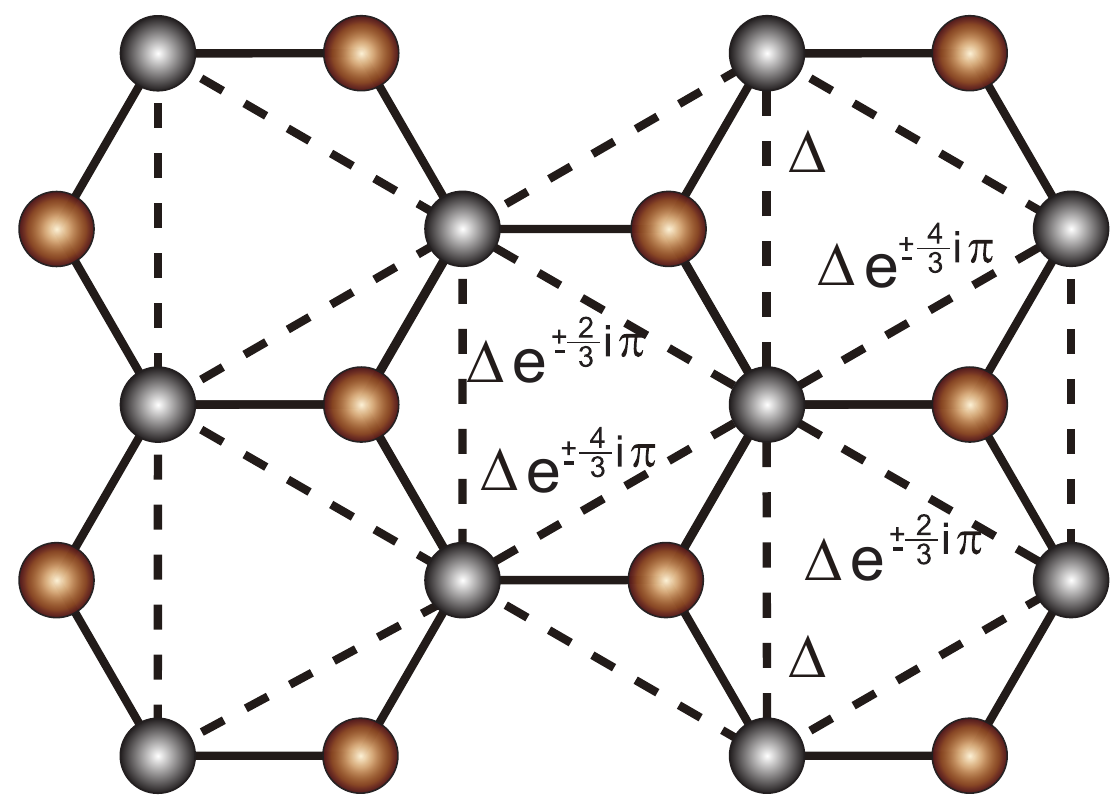}}
\caption{  The two dimensional hexagonal lattice formed by the corner-shared  trigonal bipyramidal complexes. The grey cation atoms further form a triangle lattice. The  superconducting pairing configuration  in a  $d\pm id$ pairing state is sketched.
 \label{triangle} }
\end{figure}

If we compare all cation-anion complexes,   the trigonal bipyramidal complex has   slightly lower symmetry  than the octahedral or tetrahedral complexes.  Materials with layered structures have also been formed by trigonal bipyramidal complexes, such as $YMnO_3$\cite{YMnO3-3,YMnO3-4} in which  Mn atoms in  a  Mn-O hexagonal lattice form  a triangular lattice  through  conner-shared $MnO_5$ complexes as shown in Fig.\ref{triangle}.   The d-orbitals in the trigonal bipyramidal complex are split into three groups as shown in Fig.\ref{crystal}(c).  The $d_{z^2}$ orbital has the highest energy due to its strong couplings to apical anions.  The degenerate $d_{x^2-y^2}$ and $d_{xy}$ orbitals are strongly coupled to the in-plane anions.   The degenerate $d_{xz}$ and $d_{yz}$ orbitals have the lowest energy and are only weakly coupled to anions.  Thus, one can guess that a $3d^{6}$ or $3d^{7}$ configuration  may  result in a possible band structure in which the  $d_{x^2-y^2}$ and $d_{xy}$ orbitals dominate near Fermi surfaces. If we further consider  two molecular orbitals formed by the  $d_{x^2-y^2}$ and $d_{xy}$ orbitals, one of them can strongly couple to the $d_{z^2}$. As the $d_{z^2}$ orbital has higher energy,  the coupling lowers the energy level of  this molecular orbital.  Therefore,  to form a band structure that is dominated by the pure $d_{x^2-y^2}$ and $d_{xy}$ orbitals near Fermi energy,  the $3d^{7}$ filling configuration is expected as shown in Fig.\ref{crystal}(c).

\begin{figure}[t]
\centerline{\includegraphics[height=10 cm]{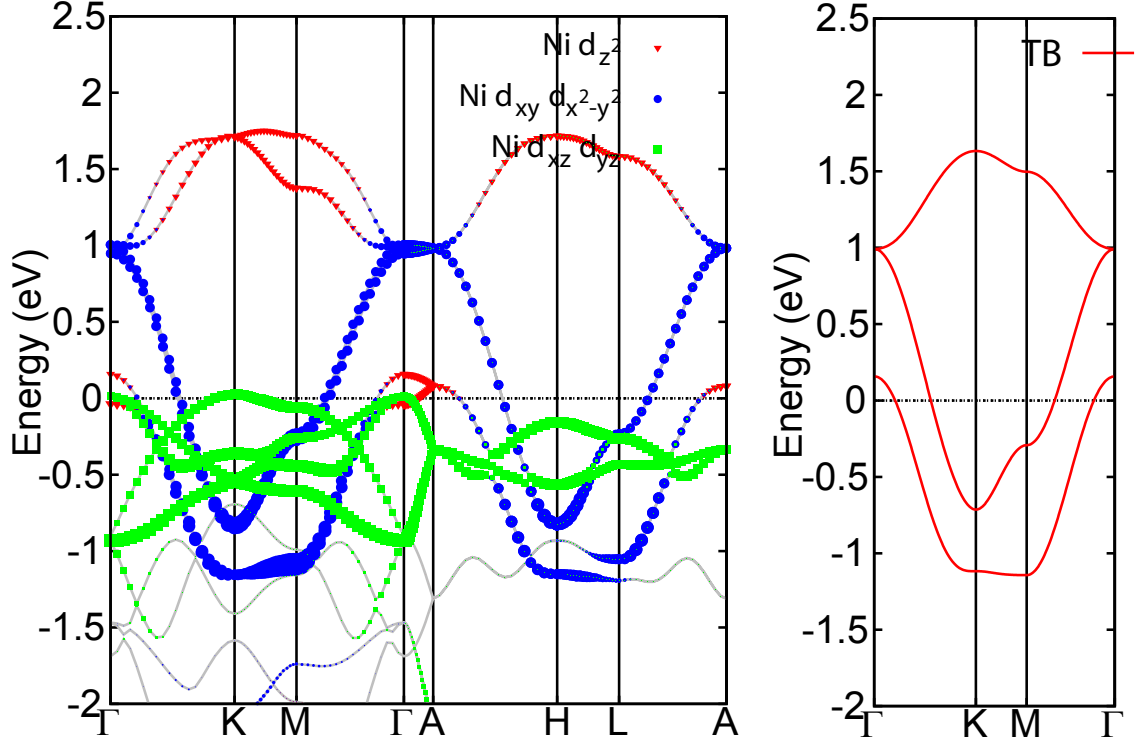}}
\caption{The band structures of $YNiO_3$ obtained from the first principle calculations and  the exacted three bands for the tight binding model.
 \label{band} }
\end{figure}

Both $Co^{2+}$ and $Ni^{3+}$ ions have a $3d^7$ filling configuration.  The $MnO_3$ layer in $YMnO_3$ is the simplest prototype layer structure  that can be formed by trigonal bipyramidal complexes without anion bonding.   Here we  focus on this prototype structure and  check whether a desired electronic structure  for high $T_c$ superconductivity  exists.    Fig.\ref{band}(a) shows the electronic band structure of $YNiO_3$.  The electronic structure   is rather quasi-two dimensional and thus can be attributed to a single $NiO_3$ layer.  In Fig.\ref{band}(a),   one band near the Fermi level, which has the largest dispersion and  will be referred   as the $\alpha$ band,  is mainly attributed to the two $d_{xy}$ and $d_{x^2-y^2}$ orbitals.   Another band that will be referred  as the $\beta$ band,  contributes a small hole pocket at $\Gamma$ point. The $\beta$ band is resulted from the bonding  between   the $d_{z^2}$ orbital  and one $d_{xy,x^2-y^2}$ molecular orbital. Near $\Gamma$ point,  the orbital character  of the $\beta$ band is mainly  $d_{z^2}$. The other band  from the anti-bonding  between  $d_{z^2}$  and  $d_{xy,x^2-y^2}$ orbitals, which will be referred as $\gamma$ band, stays at much higher energy and  is mainly attributed to the  $d_{z^2}$ orbital character.   The bands from $d_{xz}$ and $d_{yz}$ orbitals with much less dispersion are located below  the Fermi level.  Although it is possible that  these bands may contribute a small hole pockets at $K$ points, they  can be assumed to be fully occupied.   The p-orbitals of  the oxygen atoms is far below the  Fermi level. The large dispersion of the $d_{xy}$ and $d_{x^2-y^2}$ bands suggests a strong d-p hybridization. These features  are consistent with the above crystal field analysis  and suggest that the $3d^7$ filling configuration in trigonal bipyramidal complexes is indeed a possible candidate for high temperature superconductivity. Neglecting  the interlayer coupling, the electronic structure can be well described by a three-band tight binding (TB) model, including $d_{xy}$, $d_{x^2-y^2}$ and $d_{z^2}$ orbitals. Fig.\ref{band}(b) shows that  the band structure obtained from the TB model well captures  the first principle calculation results. The corresponding hopping parameters are given in Table.\ref{hopping1}. It is worth to note that the signs of intraorbital hopping parameters for the $d_{xy}$ and $d_{x^2-y^2}$ orbitals also indicate that  the hopping is caused by  the oxygen atoms.

\begin{figure}[t]
\centerline{\includegraphics[height=4.5 cm]{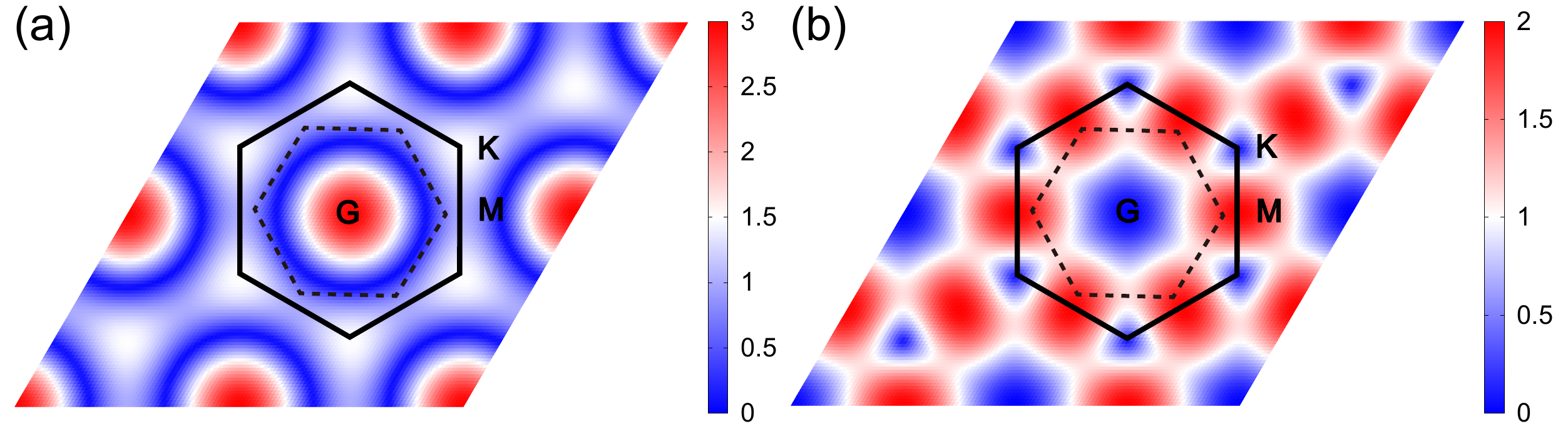}}
\caption{The overlap between Fermi surfaces of the $\alpha$ band and gap functions:  (a) $s$-wave, $cosk_y+2cos\frac{\sqrt{3}}{2}k_xcos\frac{1}{2}k_y$; (b) $d\pm id$-wave, $cosk_y-cos\frac{\sqrt{3}}{2}k_xcos\frac{1}{2}k_y\pm i \sqrt{3}sin\frac{\sqrt{3}}{2}k_x sin\frac{1}{2}k_y$. The dashed black lines represent the Fermi surfaces. The solid black lines represent the first Brillouin zone.
 \label{formfactor} }
\end{figure}

Following the second principle, the $\alpha$-band from the $d_{xy}$ and $d_{x^2-y^2}$  orbitals  can host high temperature superconductivity.  We can check whether this structure also satisfies the HDDL principle.  Near $3d^7$ filling configuration, this band is  close to half-filling.  The $\alpha$ band can be described by a simple one-dimensional effective Hubbard or t-J   models in a two-dimensional triangle lattice.  The dominant hopping parameter is the nearest neighbour (NN) hopping  and the short range magnetic superexchange  coupling is also the NN antiferromagnetic(AFM) exchange.   In the supplementary, we also show that the AFM state has significantly  lower energy than the paramagnetic state, which indicates  the existence of  the strong NN AFM exchange couplings in the parental compound $YNiO_3$. In a triangle lattice, the NN AFM exchange coupling can lead to   two types of pairing symmetries: $s$-wave  or $d\pm id$-wave\cite{Huding2012}. As the pairing  should be dominated on the NN bonds,  for the $s$-wave pairing, the   form factor of the gap function  in  the momentum space is given by  $
\Delta_s\propto cosk_y+2cos\frac{\sqrt{3}}{2}k_xcos\frac{1}{2}k_y,
$
and  similarly  for  the $d\pm id$-wave pairing, the factor is given by $\Delta_d \propto cosk_y-cos\frac{\sqrt{3}}{2}k_xcos\frac{1}{2}k_y\pm i \sqrt{3} sin\frac{\sqrt{3}}{2}k_x sin\frac{1}{2}k_y$.  Following ref.\cite{Huding2012}, we calculate the overlaps between the Fermi surfaces and the form factors.  Fig.\ref{formfactor} shows the overlaps  for the $\alpha$-band obtained in $YNiO_3$. It becomes obvious that the $d\pm id$-wave form collaborates  well with Fermi surfaces near half filling and  its' overlap with the   Fermi surfaces is much larger  than  the $s$-wave form.  Therefore, the system is a good candidate to host a high $T_c$ superconducting state with a robust $d\pm id$-wave pairing symmetry.

The $\alpha$ band is a rather robust electronic structure as long as  the two dimensional triangle lattice is maintained. Without considering the lattice instability,  we can extend the $YNiO_3$ prototype to include many possible variations  by choosing different valence anions and replacing the apical anions with different elements. In the supplementary, we provide a list of possible materials in which the $\alpha$ band stands out near the  Fermi level, including  $KNiOCl_2$, $KNiOF_2$, $BaCoOF_2$ and $KCoF_3$.   In all these prototypes, the $\alpha$ is close to the half filling  with a dispersion similar to the  one  in Fig.\ref{band}  in $YNiO_3$.    The $\beta$ and $\gamma$ bands can be tuned by changing apical anion elements.  For example,  in  the material, $KNiOCl_2$, the $\beta$ band sinks below Fermi level and has no hole pocket contribution at $\Gamma$ point. 

Similar to the octahedral complex,  the trigonal bipyramidal complex  can be flexibly crystallized into structures with  multiple triangle layers in a unit cell because of the existence of the apical anions. $YbFe_2O_4$\cite{YbFeO4} structure is one such flexible structure  with a  double-triangle-layer structure.  If we consider $YbNi_2O_4$ in the $3d^7$ configuration,  shown in the supplementary,  the $\alpha$ band  is very similar to the one in $YNiO_3$.  This proves again that $\alpha$ band is very robust and is strongly  determined by the in-plane d-p hybridization. In cuprates,  materials with multiple $Cu-O$ layers in a unit cell, such as $YBa_2Cu_3O_{7−x}$(YBCO)\cite{Wuybco},  has significantly  higher $T_c$ than the single layer materials, such as $La_{2-x}Ba_xCuO_4$(LSCO)\cite{Bednorz1986}.  The flexibility of the trigonal bipyramidal complex thus may also help these classes of materials to reach the potential maximum $T_c$.

We can estimate the possible highest $T_c$ that could be  achieved in these systems. As a rough estimation, we can compare the energy scales of the effective models with those of cuprates and iron-based superconductors.  In cuprates, the NN effective hopping parameter induced  through the d-p hybridization is about $0.43$ev\cite{Norman2003}.  In iron-based superconductors,  it is the next NN (NNN) effective hopping parameters  induced primarily by the d-p hybridization. The values of the NNN hopping parameters range from $0.15$ev to $0.25$ev\cite{Kuroki2008-prl}, depending on materials and orbitals.
Thus the energy scale in iron-based superconductors is roughly half of the energy scale in cuprates.  The highest $T_c$ in iron-based superconductors is also around the half of the value achieved in cuprates.  In the fitted TB model in Table.\ref{hopping1},  the NN hopping  is about $0.31$ev. Therefore, we expect that the highest $T_c $ here is at least comparable to those in iron-based superconductors. Namely, it should be over $50$k.      It is important to note that the above estimation is only for the possible maximum $T_c$. The superconducting transition temperature in a  superconductor, in general, is very sensitive to the detailed electronic structures, doping concentration, material quality,  possible competing orders and many other factors.

It is interesting to compare the proposed electronic structure with those of the layered sodium cobalt oxyhydrate, $NaCoO_2$, which  owns a triangular cobalt €"oxygen lattice\cite{Takada-nacoo2}.  However, the triangular cobalt lattice is built by edge-shared $CoO_6$ octahedral complexes.  The NN hopping between two Co atoms stems from the d-d direct chemical bondings. Thus,  even if the strong electron-electron correlation has been argued in this material\cite{Singh-nacoo2}, the material violates our basic principles so that it is not a candidate  for  high $T_c$ superconductivity.

We can also design the similar structure with  4d or 5d transition metal  elements as  cation atoms in the $4d^7$  or $5d^7$ filling configuration. In the supplementary, we provide the band structure of $Pd$-based materials in which $Pd^{3+}$ is  in a $4d^7$ filling configuration.   The essential $\alpha$ band is very similar to the above results.   Although the correlation effect is generally weakened in heavier transition metal systems,  the robust $\alpha$ band suggests that the proposed class of high $T_c$ superconductors may include  many series of materials.

In summary, we predict that  high $T_c$ superconductivity exists in a triangle lattice formed by  the  cation-anion trigonal bipyramidal complexes close to a $d^{7}$ filling configuration  on the cation ions.  The predicted Co/Ni based superconductors or corresponding 4d/5d transition metal based superconductors should have a robust  $d_{xy}\pm id_{x^2-y^2}$ pairing symmetry. If the prediction is verified,   together with cuprates and iron-based superconductors, it  can convincingly establishes the high $T_c$ superconducting mechanism  and also  paves a way to design  and search new unconventional high $T_c$ superconductors.

\textbf{Acknowledgement: } {We thank DL Feng for useful discussion. The work is supported by  the National Basic Research Program of China, National Natural Science Foundation of China(NSFC)
and the Strategic Priority Research Program of  Chinese Academy of Sciences. }

\begin{table}[bt]
\caption{\label{hopping1} The NN hopping parameters (in unit of eV) along  the $y$-axis in the three-orbital model. The onsite energies are $\epsilon_1=2.765$eV and $\epsilon_2=4.186$eV and the Fermi level is $E_f=3.045~eV$.}
\begin{ruledtabular}
\begin{tabular}{cccc}
  & $d_{xy}$ & $d_{x^2-y^2}$ & $d_{z^2}$  \\
 \colrule
$d_{xy}$   & 0.3147 & 0.0388 & -0.2063 \\
$d_{x^2-y^2}$ & -0.0388 & 0.1091 & 0.0678       \\
$d_{z^2}$ & 0.2063 & 0.0678 & -0.1639
\end{tabular}
\end{ruledtabular}
\end{table}

\end{document}